\documentclass[conference]{IEEEtran}
\IEEEoverridecommandlockouts
\usepackage{cite}
\usepackage{amsmath,amssymb,amsfonts}
\usepackage{algorithmic}
\usepackage{graphicx}
\usepackage{textcomp}
\usepackage[dvipsnames]{xcolor}
\def\BibTeX{{\rm B\kern-.05em{\sc i\kern-.025em b}\kern-.08em
    T\kern-.1667em\lower.7ex\hbox{E}\kern-.125emX}}

\usepackage{array}
\newcolumntype{x}[1]{>{\centering\arraybackslash\hspace{0pt}}p{#1}}

\newcommand{\probP}{\text{I\kern-0.15em P}}

\DeclareMathOperator*{\argmin}{\arg\!\min}   

\usepackage[utf8]{inputenc} 

\begin{document}

\title{Predictive Network Configuration with Hierarchical Spectral Clustering for Software Defined Vehicles}

\author{
    \IEEEauthorblockN{Pierre Laclau$^{1,2}$, Stéphane Bonnet$^1$, Bertrand Ducourthial$^1$, Xiaoting Li$^2$ and Trista Lin$^2$}

    $^1$Heudiasyc, UMR CNRS, Université de Technologie de Compiègne, France \\

    $^2$Stellantis, Vélizy-Villacoublay, France \\

    \{pierre.laclau, stephane.bonnet, bertrand.ducourthial\}@utc.fr, \{xiaoting.li, trista.lin\}@stellantis.com
}

\maketitle

\begin{abstract}

    The increasing connectivity and autonomy of vehicles has led to a growing need for dynamic and real-time adjustments to software and network configurations. Software Defined Vehicles (SDV) have emerged as a potential solution to adapt to changing user needs with continuous updates and onboard reconfigurations to offer infotainment, connected, and background services such as cooperative driving. However, network configuration management in SDVs remains a significant challenge, particularly in the context of shared Ethernet-based in-vehicle networks. Traditional worst-case static configuration methods cannot efficiently allocate network resources while ensuring Quality of Service (QoS) guarantees for each network flow within the physical topology capabilities. In this work, we propose a configuration generation methodology that addresses these limitations by dynamically switching between pre-computed offboard configurations downloaded to the vehicle. Simulation results are presented and future work is discussed.
\end{abstract}

\noindent\let\thefootnote\relax\footnote{This work was supported by Stellantis under the collaborative framework UTeam/UTC/CNRS/PCA with Heudiasyc (ANRT contract n°2021/0865).}


\section{Introduction}


In the last few years, automotive OEMs have started to develop a range of complex features such as dedicated infotainment app stores, automated driving, and Vehicle-to-Everything (V2X) services. This shift invites OEMs to continuously deliver these features through seamless Over-the-Air (OTA) updates, which contributes to longer lasting vehicles in the hope to bring the industry closer to sustainable transportation systems. As a result, future vehicles may resemble '\textit{Smartphones on Wheels}' where users can dynamically request services throughout the vehicle lifetime while OEMs continuously integrate background services such as autonomous and cooperative driving, smart grid algorithms, and more \cite{haeberleSoftwarization2020,huEnergy2018}.

To support this paradigm shift, the automotive industry is undergoing a rapid transformation toward Software Defined Vehicles (SDV) \cite{bucklSoftware2012}. Previously, Electric and Electronic (E/E) architectures were hardware-defined by integrating many single-function Electronic Control Units (ECU) into domain-centric networks. However, the increasing number of ECUs as well as more demanding network and OTA requirements are reaching the limits of current architectures \cite{bandurMaking2021}. These changes are motivating a global shift towards Zonal Oriented Architectures (ZOA), where fewer High Performance Computing (HPC) ECUs are expected to group and manage multiple heterogeneous functions. ZOA allows for a centralized and therefore cost-effective reservation of computational resources for future updates and features. In addition, Ethernet introduces service-oriented communications and increased bandwidth \cite{kimDevelopment2021}. 


This new hardware approach must be controlled by a software stack capable of applying OTA updates and dynamically switching the onboard software context to provide relevant applications. A common solution is to deploy a Service Oriented Architecture (SOA) which handles high-level communications through publish-subscribe protocols and the orchestration of services based on user requests, vehicle context, and available updates, allowing for dynamic reconfigurations of software and network resources \cite{rumezOverview2020b}. However, SOA does not address the guarantee of Quality of Service (QoS) constraints such as hard real-time latency and jitter for safety-critical flows. Hence, the onboard architecture also relies on an embedded infrastructure based on technologies such as Software Defined Networking (SDN) and Time Sensitive Networking (TSN) that can be used to handle dynamic routing and hard real-time scheduling of network resources on a per-flow basis.

Such an architecture depends on \emph{configurations} to describe the behavior of physical and virtual network components. These configurations determine how traffic shapers, priorities, buffers, routes, firewalls, sleep modes, and others are allocated to meet all mixed QoS constraints. However, generating these configurations is a complex and often NP-hard task \cite{nasrallahPerformance2019} as it requires generating a coherent global configuration across all network components despite their heterogenous implementations and parameters. For instance, some components such as 802.1Qbv for TSN rely on SAT-based solvers to generate valid configurations (e.g. TSNsched \cite{santosTSNSCHED2019}). It would thus be infeasible to generate configurations onboard due to limited resources.

We believe that current methods for determining network configurations using worst-case static approaches are no longer sufficient for SDVs due to the potential presence of thousands of diverse applications and the need for frequent updates. Therefore, it is necessary to find new approaches to manage the dynamic nature of network configurations. In this work, we propose a novel methodology for generating these configurations, which dynamically reconfigures the in-vehicle network using pre-computed offboard configurations downloaded to the vehicle. We aim to improve the flexibility and efficiency of real time embedded networks while minimizing onboard resources.

\section{Related work}


Traditionally, network configurations have been determined using static worst-case approaches, where the network is configured to handle the worst possible scenario in terms of traffic and QoS requirements for the entire vehicle lifetime \cite{quintonTypical2014,parkWorstcase2015}. There has been a considerable amount of research conducted on the generation of network configurations that varies based on the technologies and components used, such as statistical checks to validate CAN networks signal timing ranges, TSN configurators for various traffic shapers \cite{santosTSNSCHED2019}, or SDN orchestrators with dynamic routing engines \cite{halbaRobust2018}.


However, these methods assume that the input set of flows will produce a feasible configuration. Nonetheless, with the rise of SDVs and the potential for thousands of applications with diverse flow requirements, the traditional static approach becomes inadequate as it may not be able to allocate all flows into the physical in-vehicle bandwidth-limited links \cite{zhouSimulating2021}. Furthermore, with a continuously evolving offering of applications in the app store and ever-improving background services, network configurations will need to be updated regularly.

On the other hand, traditional reconfiguration methods used to apply new configurations are also limited. These methods rely on infrequent updates through OTA updates and often require reflashing ECUs and network components. This process takes time \cite{ayresContinuous2021a} and consumes onboard energy resources, making it impractical to make real-time adjustments to flow allocations. As a result, these traditional methods fail to meet the dynamic requirements of SDVs.
To address this limitation, new approaches that can handle the complexity and dynamic nature of network configurations are currently being developed. For instance, research initiatives by CoRE-RG seek to provide network reconfiguration capability at runtime for maximum orchestration flexibility \cite{hackelSecure2022a}. However, these methods only apply configurations produced by other~components.




We believe that the unique challenges posed by an environment with thousands of available services are not addressed by existing research. Our work aims to bridge this gap by leveraging existing network schedulers and dynamic reconfiguration frameworks to support (1) a large number of applications, (2) dynamic reconfigurations, and (3) minimal onboard overhead.


\section{Problem formulation}


We propose a new configuration generation methodology that allows for dynamic switching between pre-computed offboard and pre-downloaded onboard configurations. Our approach is based on the assumption that a vehicle will not require all services to be active simultaneously. We believe this to be a reasonable assumption in the context of app stores, as a large part of the available services (such as infotainment and cooperative V2X services) are functionally mutually exclusive depending on the vehicle and user contexts. 
We can thus generate smaller configurations that consider only subsets of applications that may be enabled simultaneously, thus extending the physical capabilities of the vehicle. 

\begin{figure}[t!]
    \centering
    \includegraphics[width=\linewidth]{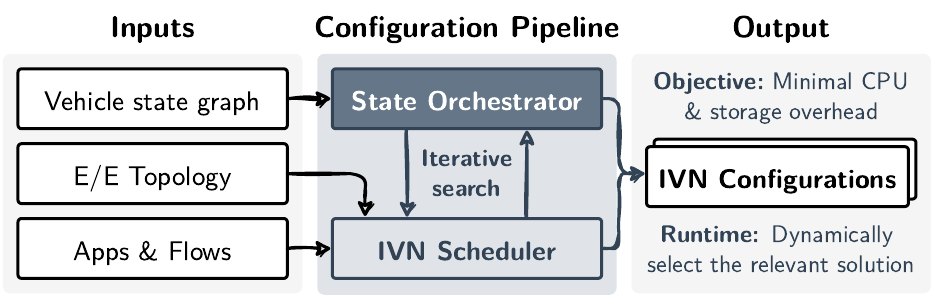}
    \caption{Block diagram of the configuration methodology. Our goal is to orchestrate the selection of network flows to produce a set of feasible configurations, using an off-the-shelf in-vehicle network scheduler for all available applications within the onboard E/E physical capacities.}
    \label{fig:arch}
\end{figure}

\begin{figure*}[t!]
    \centering
    \includegraphics[width=\linewidth]{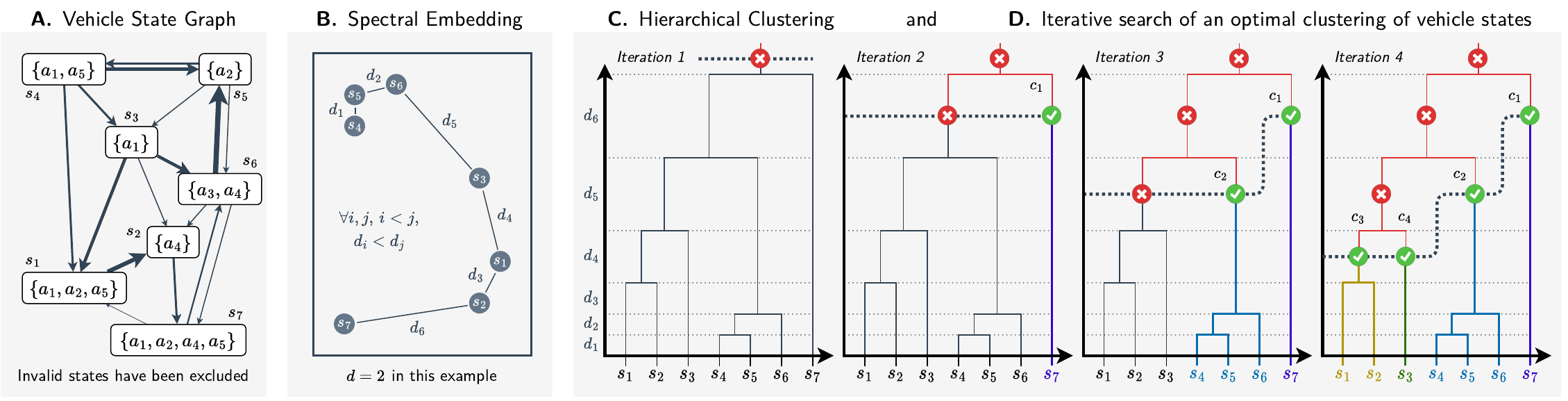}
    \caption{Illustration of our predictive configuration methodology. \textbf{(A)} represents a vehicle state graph $G$ modelled as Markov chain with nodes defined as sets of active applications and weighted edges representing reconfiguration probabilities. \textbf{(B)} shows the 2-dimensional spectral embedding of $G$, and \textbf{(C)} points to the generated tree structure called dendrogram using hierarchical clustering. \textbf{(D)} shows our iterative search algorithm, which explores the tree from the root downwards and calls the network scheduler at each branch to get its validity (green/red icons) until the final set of configurations $C=\{c_1, ..., c_k\}$ is found.}
    \label{fig:algo}
\end{figure*}

Let $A = \{a_1, ..., a_n\}$ be the set of applications available in the app store, including background services, along with $S = \{s_1, ..., s_m\}$ a finite set of states where each state $s_i \subset A$ is a set of applications that could be activated concurrently. Each state may require a residual free bandwidth for unallocated \emph{best-effort} streams in advance, and it is assumed that all application dependencies are respected. For each pair of states $s_i$ and $s_j$, let $\probP(i, j)$ be the probability of transition from state $i$ to state $j$ and $P\!_S = \{\probP(i, j), \,\, s_i, s_j \in S\}$ the transition probability matrix linking all states of $S$. Therefore, we can define the vehicle state graph $G = \{S, P\!_S\}$, illustrated with an example in Figure \ref{fig:algo}A, as a strongly connected Markov chain which represents the different combinations of active applications that the vehicle may encounter at runtime. The state of the vehicle thus evolves as a random-walk on $G$.


The problem is to find the set of feasible configurations $C$, each generated by an architecture-dependent global network scheduler, that meets industry-relevant priorities defined below.


First, we minimize computing overhead by reducing the frequency of transitions between configurations. This optimizes resources dedicated to functional features, reduces hardware costs and energy usage, and minimizes user wait times when using traditional re-flashing reconfiguration methods (e.g. for legacy ECUs). 
Secondly, we mitigate the amount of data transmitted and stored onboard by minimizing the number of configurations that need to be generated. While the car will likely often have WiFi available while parking, this may not be true all the time which would require paid mobile data usage.

Note that these two objectives are only sufficient to guarantee safe reconfigurations while parked, meaning no active flows remain active during transitions. While driving, some packets from previously active applications might still be in transit inside the network, which would potentially lead to lost packets as configurations are generated independently with separated buffer usage estimates. In this work, we focus on orchestrating the generation of independent configurations by building on top of existing network schedulers for parked reconfigurations, as current schedulers do not support the generation of multiple coherent configurations. 

Finally, we minimize the number of calls to the network scheduler as it might take a considerable amount of time to check the validity of a particular set of flows. The configurations depend on the E/E physical topology which includes the set of ECUs available in the vehicle $E$, as well as the list of all potential flows $F$ required by applications as inputs for the network scheduler. Hence, each application is defined as $a_i=\{e, f\}$ with $e \in E$ the predetermined host ECU for its execution and $f \subset F$ the set of QoS-dependent flows required. This serves as the input data for the global network scheduler. The diversity of parameters and flow types can vary depending on the capabilities of the selected scheduler, which can be considered as as a \emph{black box} in this work. 

In an industrial context, the vehicle state graph $G$ can be generated based on expert system modelling, vehicle simulation, and real-world observation feedback. However, we assume that $G$ might contain thousands of states, some of which could be subsets of other states. Hence, our objective is to process this input graph such that the final set of pre-computed configurations remains minimal.
Therefore, finding $C$ can be done by solving the following optimization problem:

\begin{equation}\label{eqn:optprob}
\argmin_C g(C)
    \quad \text{with} \quad g(C) = k + W + h 
\end{equation}
$$
    \text{such as}
    \quad k = |C|
    \quad \text{and}
    \quad W = \sum_{s_i, s_j \in C} P\!_{C}(i,j)
$$

\noindent where $C$ is the final set of generated configurations, $k$ is the number of configurations, $W$ is the sum of transition probabilities between configurations with $P\!_C$ the transition matrix constructed from the subgraph of $G$ clustered using $C$, and $h$ is the number of calls to the global network scheduler.
\section{Methodology}

The purpose of the \emph{State orchestrator}, illustrated in Figure~\ref{fig:arch}, is to orchestrate the selection of flow sets to achieve the previous optimization goal. 
Our methodology, illustrated in Figure \ref{fig:algo}, consists of the following steps.

\begin{figure*}[t!]
    \centering
    \includegraphics[width=\linewidth]{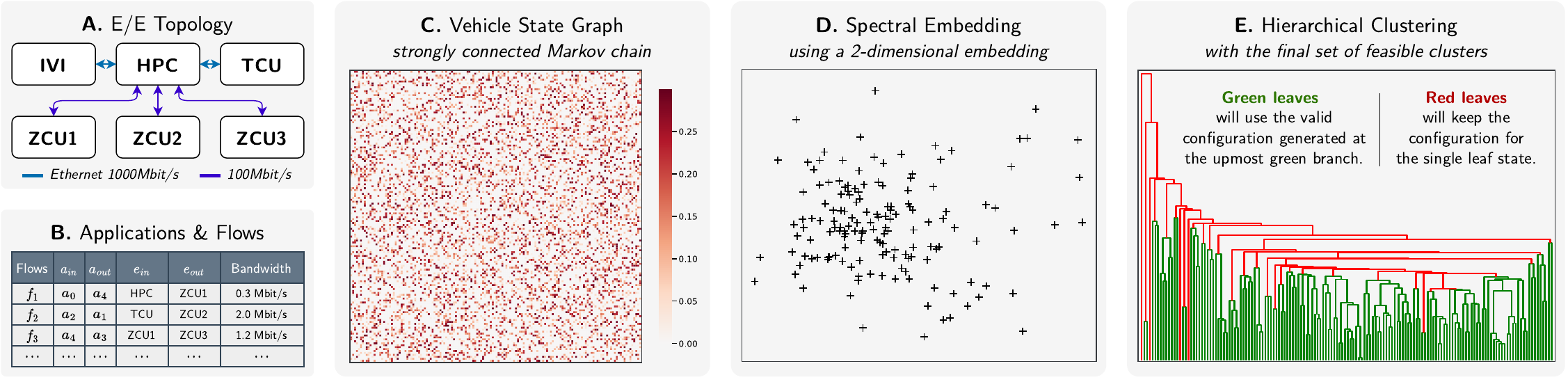}
    \caption{Simulation results. We generate a simple scenario with \textbf{(A)} a zonal-oriented E/E architecture topology, \textbf{(B)} a random set of flows with parameters supported by the scheduler model, and \textbf{(C)} a heatmap that represents the $P\!_S$ transition matrix of a random strongly connected Markov chain $G$. \textbf{(D)} shows the 2-dimensional spectral embedding of $G$, and \textbf{(E)} presents the generated dendrogram with the final set of iteratively searched configuration clusters. 
    }
    \label{fig:results}
\end{figure*}

\textbf{Initialization (Figure \ref{fig:algo}A).} \quad First, it is possible to filter $G$ by running the scheduler on each state to remove unfeasible states. Similarly, it is possible to filter out states already included in others based on the remaining bandwidth and energy consumption goals of the vehicle.

\textbf{Projection (Figure \ref{fig:algo}B).} \quad Next, we apply the \textit{spectral embedding} method \cite{sanchez-garciaHierarchical2014} where $d$ is the destination dimension to be calibrated in the industrialization phase such that $2 \leq d \ll |S|$. Thus, the relative coordinates of each state in the embedded space correspond to their probabilistic relations. This has the effect of moving states with high transition frequencies closer together. Since $G$ is directed, we use the asymmetric Laplacian matrix as the similarity matrix between nodes to account for irregular probabilities \cite{zhengSpectral2015}.

\textbf{Hierarchical Clustering (Figure \ref{fig:algo}C).} \quad We then apply the \textit{agglomerative hierarchical clustering} method \cite{sanchez-garciaHierarchical2014} using the embedded space by merging the states in pairs following the increasing order of Euclidean distance (Y axis in Figure \ref{fig:algo}C) between states, which groups states with high transition probabilities first. This results in a tree structure according to their probabilistic relationships. At this point, we obtain a \emph{dendrogram} where each branch represents a candidate cluster of states. Each branch can be seen as a new state coming from the merging of the applications contained in the children~states.

Each time two states $s_i$ and $s_j$ are clustered, the following occurs: first, the number of clusters $k$ decreases by $1$, which invites us to select the highest branches in the tree. Second, the transitions $P\!_S\,_{ij}$ or $P\!_S\,_{ji}$ become null as both states are merged into $s_i \cup s_j$ and the vehicle does not need to reconfigure its network as long as the active set of applications remains within this new set. Finally, it is necessary to execute the scheduler to know the validity of each branch.

\textbf{Iterative search (Figure \ref{fig:algo}D).} \quad It follows from the three preceding remarks that $g(C)$ can only increase as we explore the dendrogram starting at the root downward by running the scheduler at each branch encountered. The algorithm thus consists in finding the branches of maximum height in the dendrogram producing valid configurations by exploring the topology from top to bottom. Since the algorithm stops at the first combination of valid branches found and $g(C)$ increases during the exploration, the solution corresponds to the minimum of $g$ while constrained to return a set of valid configurations. In the end, all states $s_i$ obtain a valid configuration generated at one of the parent branches.

The algorithm starts with the root of the topology $c_r$ which corresponds to the ideal solution that minimizes all objectives. It is also equivalent to the worst case static scheduling methods found in the state of the art. We run the network scheduler on this single merged state and return $C = \{c_r\}$ if the scheduler generates a valid configuration. Otherwise, we repeat the procedure independently for all children of the node until the first valid branches are found. This later case is more probable in the context of thousands of available applications. Thus, we obtain $C = \{c_1 , ..., c_k\}$ solution of the optimization problem posed in Equation~\ref{eqn:optprob}.

The overall algorithm is of polynomial complexity. The complexity of the setup stage (corresponding to steps A, B, and C in Figure \ref{fig:algo}) is $\mathcal{O}(n^2)$ with $n = |S|$ (e.g. see \cite{sanchez-garciaHierarchical2014}). This phase only needs to be performed once with a given set of applications. Then, the iterative search (Figure \ref{fig:algo}D) is reduced to $\mathcal{O}(n)$ since at most the $2n - 1$ nodes of~the~cluster tree topology need to be explored, which allows for an efficient generation of configuration sets for multiple physical vehicle topologies using a single dendrogram. Note that each scheduler call for a candidate branch $c_i$ has a complexity that varies with the chosen model, such as $\mathcal{O}(|c_i|^2)$ for TSNsched \cite{santosTSNSCHED2019}.



\section{Simulation setup}

To test the performance of our approach, we designed a simulation setup with a simple but representative scheduler model using randomly generated inputs. First, the E/E physical topology shown in Figure \ref{fig:results}A follows a standard zonal-oriented star topology connected with Ethernet links of different datarates. Then, we generate a random dependency graph of $n=500$ applications to represent an app store as a binomial directed graph.
Each application is randomly assigned to one of the ECUs, which we assume is provided by an external onboard service orchestrator.

We selected a simple scheduler model that considers a unique \emph{bandwidth} parameter and checks if the sum of the active flows in each link does not exceed its physical capacity. Then, we generate a random number of $1$ to $5$ flows for each edge in the dependency graph, with each being assigned a randomized bandwidth requirement between $0.1$ and $5$~Mbit/s
(see Figure \ref{fig:results}B). Additional constraints such as latency and jitter could be added when using a more complex~scheduler.

These inputs are sufficient to run traditional worst case methods. To compare these methods with our dynamic approach, we also generate the random strongly connected Markov chain $G$ shown in Figure \ref{fig:results}C. This is done by taking the largest strongly connected subgraph of a random directed graph with $m=500$, which produces a graph of $m=433$ states. We then produce a random combination of active applications such that $s \subseteq A$ by sampling a random set of $5$ to $500$ applications and adding their dependencies. We also replaced existing states until the union set of all states contained all applications in $A$ to produce a coherent input~set.

Then, we filter $G$ by removing states with unfeasible configurations for this particular E/E topology. Note that these invalid states may become valid for higher-end topologies. This methodology produces a final graph $G$ of size $m=150$. Edges are valued using random transition weights. Our algorithm does not assume any similarity between the sets of active apps between neighbor states in $G$, since its goal is to minimize the frequency of reconfigurations.
Finally, we define $r_A$ as the ratio of applications in $A$ that have been allocated in at least one valid configuration for each scenario.

We study three simulation scenarios. Firstly, we attempt to schedule all flows at once by calling the scheduler once, which is equivalent to the worst case strategies (\emph{worst-case} scenario). If the scheduler does not return a valid configuration, we sample random combinations of applications using the same methodology to generate $S$ in order to estimate the performance of this strategy in an app store environment. Secondly, we generate one configuration for each state $s_i$ in the state graph (\emph{unfiltered} scenario). Lastly, we run our clustering strategy to reduce the number of final states (\emph{reduced} scenario).

To evaluate the performance of each scenario, we compare each part of the objective function $g$, namely the number of produced configurations $k$, the sum of transition probabilities between states $W$, and the total number of scheduler calls $h$.

\section{Results}

The simulation results are summarized in Table \ref{tab:results}. They indicate that the worst-case static approach is insufficient for allocating all flows in a single configuration. When sampling $1000$ random feasible application combinations, the maximum amount of apps that can be allocated to a single configuration using the worst-case approach was found to be only $28\%$ with a mean of $13\%$. Even though $g(c_r)$ is minimal, this implies that the worst-case approach cannot guarantee an efficient allocation of all flows in a single configuration.

On the other hand, generating one configuration for each state $s \in S$ in the initial state graph resulted in $100\%$ of valid configurations as expected by design. However, this approach generated a large number of configurations $k=150$, which makes it impractical to be used in real-world scenarios due to the increased complexity and storage requirements. This solution also requires frequent onboard reconfigurations.

The proposed approach achieved better performance, generating $39$ configurations while maintaining $100\%$ valid configurations with transition probabilities reduced by $87.9\%$ and scheduler calls by $62\%$. This suggests that our approach is efficient in allocating all flows with valid configurations, which makes it more practical for real-world deployment.

Our results demonstrate the effectiveness of our proposed methodology in reducing the number of configurations required while achieving a $100\%$ success rate in valid configurations. This approach can be used in real-world scenarios where the number of ECUs and applications can be much larger, and hence the proposed methodology can provide an effective solution for dynamic configuration allocation.

Our results highlight the importance of a dynamic approach to configuration allocation, and the proposed approach can help improve the efficiency and scalability of automotive networks while ensuring a high level of QoS for all flows.

\begin{table}[t!]
    \caption{Evaluation metrics for the simulation results
    }
    \centering
    \begin{tabular}{c||c|c|c||c}
         \multicolumn{1}{x{0.3cm}||}{} & \multicolumn{1}{x{1.4cm}|}{Worst case \newline $c_r$} & \multicolumn{1}{x{1.4cm}|}{Unfiltered \newline $S$} & \multicolumn{1}{x{1.4cm}||}{Reduced \newline $C$}  & \multicolumn{1}{x{1cm}}{\hspace{1mm} Gains \newline $S \rightarrow C$}\\
        \hline
        $r_A$    & {\color{Red}$13\%$ (mean)
        } & {\color{ForestGreen}$100\%$} & {\color{ForestGreen}$100\%$} & {\color{ForestGreen}$0\%$}      \\
        $k$                       & $1$                          & $150$                  & $39$                      & {\color{ForestGreen}$74\%$}\\
        $W$                       & $0$                          & $3411.5$               & $413.7$                   & {\color{ForestGreen}$87.9\%$}\\
        $h$                       & $1$                          & $150$                  & $57$                      & {\color{ForestGreen}$62\%$}\\
        
    \end{tabular}
    \label{tab:results}
\end{table}

\section{Conclusion}

We have presented a methodology for generating a set of offline network configurations to cover the dynamic use cases of vehicles. The configurations are generated with the aim to minimize the frequency of reconfigurations in the vehicle and storage requirements while reducing the number of network scheduler calls. Finally, the vehicle dynamically selects the correct pre-downloaded configuration on context changes.

This approach applies classical clustering strategies in an industrial environment to enable a large number of applications despite limited onboard hardware capabilities. Additionally, it is possible to optimize the offboard execution times in an industrial context by starting the exploration at an intermediate level in the dendrogram or by skipping multiple descendants at once, depending on calibration. Furthermore, the branches are independent and can therefore be evaluated in parallel.

This scenario can be particularly useful in the context of shared vehicles, where the vehicle reconfigures itself before a user reservation period based on their preferences, profile, and subscriptions. Moreover, the same state graph $G$ can be used for several vehicle models with different topologies. However, one major limitation remains: this algorithm is independent from the network scheduler, which makes it impossible to construct multiple coherent configurations based on QoS guarantees to ensure safe reconfigurations on running vehicles. 

Future work will aim to (1) test the network performance of our solution in a realistic automotive environment using more complex schedulers such as TSNsched \cite{santosTSNSCHED2019} along with realistic input applications and flows, (2) improve current schedulers to consider transition safety and enable reconfigurations while driving, (3) extend the current architecture with an onboard configuration generator based on heuristics for non-critical flows, and (4) study the impacts of different configuration orchestration strategies on onboard energy consumption.






\bibliographystyle{IEEEtran}
\bibliography{references}

\end{document}